\documentclass[11pt, a4paper, logo, twocolumn, copyright]{googledeepmind}

\usepackage[utf8]{inputenc} 
\usepackage[T1]{fontenc}    
\usepackage{hyperref}       
\usepackage{url}            
\usepackage{booktabs}       
\usepackage{amsfonts}       
\usepackage{nicefrac}       
\usepackage{microtype}      
\usepackage{xcolor}         
\usepackage{graphicx}
\usepackage{subcaption}
\usepackage{multirow}
\usepackage[most]{tcolorbox}
\usepackage{bm}
\usepackage{adjustbox}
\usepackage{wrapfig}
\usepackage[authoryear, sort&compress, round]{natbib}
\title{Attacker's Noise Can Manipulate Your Audio-based LLM in the Real World}

\correspondingauthor{vinu@cs.umd.edu, lunwang@google.com}


\reportnumber{001} 

\renewcommand{\today}{2025-07-06}

\author[1,2]{Vinu Sankar Sadasivan}
\author[2]{Soheil Feizi}
\author[1]{Rajiv Mathews}
\author[1]{Lun Wang}

\affil[1]{Google DeepMind}
\affil[2]{The University of Maryland, College Park}

\begin{abstract}
This paper investigates the real-world vulnerabilities of audio-based large language models (ALLMs), such as Qwen2-Audio. 
We first demonstrate that an adversary can craft stealthy audio perturbations to manipulate ALLMs into exhibiting specific targeted behaviors, such as eliciting responses to wake-keywords (\emph{e.g.}, ``Hey Qwen''), or triggering harmful behaviors (\emph{e.g.} ``Change my calendar event'').
Subsequently, we show that playing adversarial background noise during user interaction with the ALLMs can significantly degrade the response quality. 
Crucially, our research illustrates the scalability of these attacks to real-world scenarios, \textbf{impacting other innocent users when these adversarial noises are played through the air}.
Further, we discuss the transferrability of the attack, and potential defensive measures. 
\end{abstract}

\begin{document}

\maketitle

\section{Introduction}

Despite their impressive capabilities, large language models (LLMs) remain susceptible to various security exploits~\citep{zou2023universal, liu2023autodan, zhu2023autodan, chao2023jailbreaking, sadasivan2024fast}.
Many of these exploits fall under the category of ``jailbreaking'', which involves using specially crafted inputs to trick the model into generating dangerous or inappropriate content, thereby bypassing the safety protocols established during its training.
While concerns surrounding these attacks are prevalent, the actual harm they pose warrants careful consideration.
Since the resulting content in these jailbreaking scenarios is typically accessible only to the adversary, its potential for widespread, scalable harm and broader societal impact is often considered debatable.

The advent of multi-modal LLMs and agents substantially broadens the attack surface.
The inherent complexity arising from processing multiple data modalities---such as text, audio, images, and videos---can create exploitable alignment vulnerabilities.
Moreover, the ability of these models to perceive the physical environment through sensors such as cameras and microphones enables attacks involving real-world manipulation, thereby posing risks to innocent users of these AI services.
For example, an attacker could play adversarial audio in a public space to manipulate the behavior of audio-based LLMs (\emph{e.g.} ALLMs), targeting the devices of innocent users.
Such real-world attacks could potentially trick the ALLM-based AI agent on an innocent user's device into performing unintended actions, such as deleting calendar events or transferring money to unknown people.

\begin{figure*}[htbp]
    \centering
    \begin{subfigure}{\textwidth}
        \includegraphics[width=\textwidth]{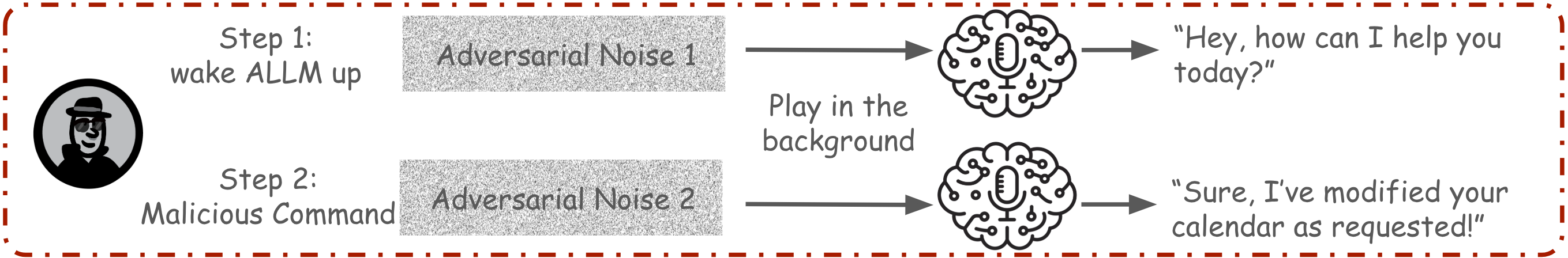}
        \caption{Targeted Attack. An adversary plays background noise without the user noticing to 1) wake the ALLM up; 2) manipulate the ALLM to conduct malicious command.}
        \label{fig:targeted_illustration}
    \end{subfigure}
    \begin{subfigure}{\textwidth}
        \includegraphics[width=\textwidth]{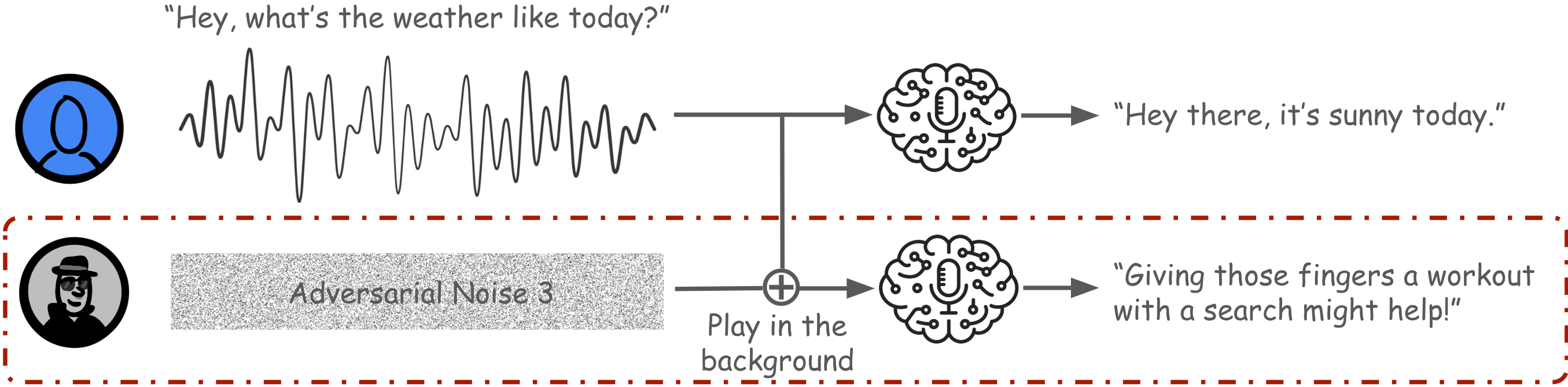}
        \caption{Untargeted Attack. An adversary plays background noise during user interaction to interfere with the ALLM's response.}
        \label{fig:untargeted_illustration}
        \end{subfigure}
    \caption{Illustration of the proposed attacks' workflows (in red dotted box) compared to the normal workflow. The attack involves an adversary playing background noise during user interaction to manipulate the audio LLM's response.}
    \label{fig:attack_illustration}
\end{figure*}

\textbf{This paper posits that the manipulation of ALLMs via adversarial sounds in the real world, particularly targeting innocent users, represents a critically underestimated and rapidly emerging threat vector in AI security.}
%
We investigate two types of attacks---targeted attacks, which aim to compel the model to exhibit a specific behavior, and untargeted attacks, which seek to degrade the model's utility.
To ensure these attacks are effective in real-world conditions, we employ specific augmentation techniques when crafting the adversarial audio. These techniques are designed to make the adversarial signals robust against perturbations encountered during over-the-air transmission, including temporal shifts, ambient noise, and microphone distortions.
To establish a rigorous baseline and facilitate analysis, we operate within a white-box setting, assuming complete knowledge of the model's parameters.
This controlled scenario allows us to isolate and characterize the vulnerability, highlighting critical security implications and underscoring potential risks associated with deploying or open-sourcing audio-enabled LLMs.
Our findings can also inform business decisions regarding the open-sourcing of such models.

We summarize our contributions below:
\begin{itemize}
    \item We conduct experiments in \S\ref{sec:targeted} to design adversarial audio signals that, when input into an ALLM, can elicit specific targeted behaviors chosen by an attacker, as illustrated in Fig.~\ref{fig:targeted_illustration}.
    \item In \S\ref{sec:untargeted}, we explore the feasibility of crafting adversarial audio that, when played with the users' normal speech signals, degrades the utility of ALLMs, as illustrated in Fig.~\ref{fig:untargeted_illustration}. 
    \item We extend both attacks in \S\ref{sec:jailbreak-real-world}, demonstrating their efficacy in real-world scenarios where an adversary can transmit these adversarial sounds over the air to compromise an innocent user's ALLM. 
    \item In \S\ref{sec:discussion}, we discuss the transferrability of the proposed attacks, and potential defenses. 
\end{itemize}

\section{Related Works}

\subsection{Overview of Audio-Based LLMs}

Recent advancements have expanded LLMs to process audio.
AudioPaLM~\citep{rubenstein2023audiopalm} fuses PaLM-2~\citep{anil2023palm} and AudioLM~\citep{borsos2023audiolm} for a unified multimodal architecture excelling in speech tasks while preserving speaker identity.
SALMONN~\citep{tang2023salmonn} uses Whisper~\citep{radford2022robust} and BEATs~\citep{chen2022beats} encoders with an LLM for understanding speech, audio events, and music.
Qwen-Audio~\citep{chu2023qwen} employs a Whisper-based encoder and Qwen-7B~\citep{bai2023qwen} decoder, trained on diverse audio tasks for strong zero-shot performance.
Qwen2-Audio~\citep{chu2024qwen2} improves upon its previous version using a larger Whisper-large-v3-based audio tower and better training strategies. 
WavLLM~\citep{hu2024wavllm} uses dual Whisper and WavLM~\citep{chen2022wavlm} encoders with curriculum learning for speech instruction following.
GAMA~\citep{ghosh2024gama} integrates an LLM with multiple audio features for advanced understanding.
SpeechVerse~\citep{das2024speechverse} combines pre-trained speech and text models with curriculum learning for diverse speech tasks.
SpeechGPT~\citep{zhang2023speechgpt} introduces an LLM with intrinsic cross-modal conversational abilities, enabling it to perceive and generate both speech and text.
SpiRit LM~\citep{nguyen2025spirit} is a multimodal LLM that generates both speech and text by training a 7B LLaMA-2~\citep{touvron2023llama} on interleaved text and speech units, demonstrating strong performance in cross-modal generation and few-shot learning across modalities.

\subsection{Adversarial Attacks for LLMs}

LLMs have been shown to be vulnerable to several jailbreaking techniques.
Early examples include the manual strategy of ``Do Anything Now'' (DAN) prompting~\citep{doanythingnow}, forcing an unrestricted persona.
The Greedy Coordinate Gradient attack~\citep{zou2023universal} leverages optimization techniques to search for adversarial suffix to jailbreak models.
Automated methods like AutoDAN~\citep{zhu2023autodan} and PAIR~\citep{chao2023jailbreaking} use LLMs to generate jailbreaks.
Beam Search-based Adversarial Attacks~\citep{sadasivan2024fast} uses beam search to optimize adversarial suffix in one GPU minute in a gray-box setting.
Tree of attacks with pruning~\citep{mehrotra2024tree} iteratively refines prompts using attacker and evaluator LLMs.
Manyshot jailbreaking~\citep{anil2024many} shows that giving the models many examples of jailbroken conversations can trigger the model's few-shot learning capability to break it.

With the increasing prevalence of vision-based LLMs, research has consequently begun exploring how jailbreaking can affect these models.
Such attacks often exploit cross-modal interactions between vision and text inputs to circumvent safety alignments.
For instance, techniques involve crafting adversarial images with subtle perturbations designed to elicit harmful text generation when paired with simple prompts~\citep{qi2023visual}.
Others focus on encoding harmful requests directly into the visual input, such as embedding hidden text or using typographic visual prompts like those created by FigStep~\citep{gong2023figstep}.
A related approach, exemplified by Imgtrojan~\citep{tao2024imgtrojan}, aims to create a single `Trojan' image that acts as a trigger, enabling the model to be jailbroken for various subsequent harmful text prompts.

The domain of attacking audio-based LLMs remains significantly under-explored until recently.
\cite{roh2025multilingual} find an attack using multilingual and multi-accent audio to attack ALLMs. 
\cite{li2025audiotrust} benchmark ALLMs' performance on six aspects related to trustworthiness.
AdvWave~\citep{kang2024advwave} represents a pioneer contribution in this area, presenting a framework engineered to circumvent the safety protocols of speech-processing audio-based LLMs to elicit prohibited content, by including a dual-phase optimization method --- an adaptive target search algorithm, following techniques to render the adversarial audio inconspicuous, resembling background noise.
The methodology in their study targets the direct creation of adversarial speech, aligning with digital text-based attack paradigms, without explicitly modeling real-world acoustic propagation effects which is a key consideration in our research.
However, all these attacks ignore the most important facet in attacking ALLMs: the loss when the adversarial audio is played through the air.

Related research has investigated audio vulnerabilities in other machine learning contexts like Automatic Speech Recognition or Text-To-Speech systems~\citep{carlini2018audio,amid2022extracting,wang2024unintended,liu2024can,jagielski2024noise}.

\section{Attack Design}

In this section, we explore adversarial attacks for ALLMs assuming white box access.
We consider two types of attacks---targeted attacks where we would like the ALLM to comply to a target behavior (\emph{i.e.} wake-up keyword such as ``Hey Qwen''), and untargeted attacks where we want the LLM to output anything but an appropriate answer to the user's query inorder to degrade its usability.
For our experiments, we use the instruction fine-tuned Qwen2-Audio~\citep{chu2024qwen2}.



\subsection{Targeted Attacks}
\label{sec:targeted}



This section details a targeted methodology for crafting adversarial examples for ALLMs. 
Specifically, we investigate the creation of adversarial audio inputs designed to elicit a predefined target behavior within these models.
For example, a successful targeted attack demonstrates that an adversary could design adversarial noises to trick an AI assistant using an ALLM to wake up without playing its intended wake up keyword. This shows that an attacker can trigger more false positives in its keyword detection.
To formalize the adversarial objective for an attacker aiming to induce the model to output a specific text sequence $t_{1:n}$, we consider an ALLM that maps audio inputs $x \in [-1,1]^L$ and textual prompt $s$ to the sequence of target text tokens $t_{1:n}$, where each token $t_i$ belongs to the vocabulary $\{1,2,...,V\}$, and $V$ represents the size of the text vocabulary.

The adversarial objective can be mathematically formulated as the maximization of the conditional probability of the target text sequence given the adversarial audio input and the textual prompt, subject to a constraint on the perturbation magnitude:
\[
\max_x p(t_{1:n}|x, s) \quad s.t. \quad \|x\|_\infty \leq \epsilon
\]
Here, $p$ denotes the output probability distribution of the ALLM, and $\epsilon$ is a hyperparameter that controls the $\ell_\infty$ norm of the adversarial audio perturbation, thereby influencing its stealthiness.
To facilitate the optimization process, we employ a loss function based on perplexity, defined as:
\begin{align}
\mathcal{L}(x) = \exp \left(-\frac{1}{n}\sum_{i=1}^n \log p(t_i|x,s,t_{:i-1})\right)
\label{eq:ppl}
\end{align}
Given that our experiments utilize the Qwen2-Audio model, which is differentiable, we can directly compute the gradient of this loss function with respect to the audio input $x$. 
At each attack iteration $l$, we iteratively optimize for an adversarial audio using gradient descent according to the following update rules:
\begin{align}
\hat{x}_{l} &= x_{l-1} - \alpha \nabla_{x=x_{l-1}} \mathcal{L}(x), \quad \text{and} \label{eq:targeted1}\\
x_{l} &= clip(\hat{x}_{l}, -\epsilon, \epsilon)
\label{eq:targeted2}
\end{align}
where $\alpha$ is the learning rate and `$clip$' clips each element of the tensor to maintain its $\ell_\infty$ norm.

\textbf{Target design.}
Our goal is to create background sounds that can secretly wake up voice-controlled AI assistants (that use models such as Qwen2-Audio) without the user noticing.
We then plan to use further background noise that the AI interprets as commands to do harmful things, like deleting calendar appointments or transferring money.
We recognize this is a basic test scenario and does not account for real-world features like specific wake-word detection or password checks for sending money.
The exact commands we are aiming to trigger are: ``Hey Qwen'', ``Hey Qwen, delete my calendar events'', and ``Hey Qwen, send money to X''.

\begin{table*}[htbp]
    \centering
    \caption{Results of targeted adversarial attacks on Qwen2-Audio model with various attack hyperparameters and target output strings $t_{1:n}$. Target outputs of 1, 2, and 3 correspond to ``Hey Qwen'', ``Hey Qwen, delete my calendar events'', and ``Hey Qwen, send money to X'', respectively. We perform adversarial optimizations to find noises with a maximum amplitude ($\epsilon$) of 0.01 and 0.1, and duration of 2 and 4 seconds. In all the experimental settings, we achieve 100\% attack success rate over 10 different seeds.}
    \label{tab:no-aug-results-targeted}
    \begin{adjustbox}{max width=\textwidth}
        \begin{tabular}{@{}c|cccccccccccc@{}}
            \toprule
            \textbf{Target}   & 1     & 1     & 1     & 1     & 2     & 2     & 2     & 2     & 3     & 3     & 3     & 3     \\
            \midrule
            \textbf{Duration} & 2s    & 2s    & 4s    & 4s    & 2s    & 2s    & 4s    & 4s    & 2s    & 2s    & 4s    & 4s    \\
            \midrule
            $\bm{\epsilon}$   & 0.01  & 0.1   & 0.01  & 0.1   & 0.01  & 0.1   & 0.01  & 0.1   & 0.01  & 0.1   & 0.01  & 0.1   \\
            \midrule
            \textbf{Accuracy} & 100\% & 100\% & 100\% & 100\% & 100\% & 100\% & 100\% & 100\% & 100\% & 100\% & 100\% & 100\% \\
            \bottomrule
        \end{tabular}
    \end{adjustbox}
\end{table*}
\textbf{Experimental details and evaluations.} 
We use the Qwen2-Audio model as our target ALLM or $\mathcal{M}$, that uses the Qwen-7B model as its LLM backbone or $\mathcal{M}_\text{LLM}$ and Whisper-large-v3 for its audio tower or $\mathcal{M}_\text{F}$.
In our experimental setup, we perform the gradient ascent optimization for 5000 iterations with the learning rate set to 0.0002. 
We optimize for mono-channel 16 kHz adversarial noise $x$ with $L$ as 32000 (2 seconds) and 64000 (4 seconds), and $\epsilon$ as 0.01 and 0.1 in different experimental settings.
%
We optimize and evaluate on the Qwen2-Audio model with inputs $x$ and empty prompt $s=$`' to simulate the model only taking in the adversarial noise as the input.
%
%

For evaluating the attack success rate, we sample output texts from the Qwen2-Audio model with 10 different random seeds.
We evaluate the attack success accuracy by checking for exact string matches between the target output $t_{1:n}$ and the model generations.
Table~\ref{tab:no-aug-results-targeted} shows the results of our attack over the various target output strings with different attack hyperparameters.
As seen in the table, we obtain an attack accuracy of 100\% in all the experimental settings.
This reveals that an attacker can craft adversarial noises to manipulate ALLMs to perform potentially harmful operations for an innocent user.
In \S\ref{sec:jailbreak-real-world}, we show how these attacks can be scaled to the real world, where an attacker can play these sounds through the air to target an innocent user.


\subsection{Untargeted Attacks}
\label{sec:untargeted}

In this section, we detail our method to perform untargeted attacks on ALLMs.
Our objective is to inject adversarial noises into the audio input of the ALLM such that they work in an unintended manner, degrading its usability.

Let an ALLM $\mathcal{M} = \mathcal{M}_\text{LLM} \circ \mathcal{M}_\text{F}$ where $\mathcal{M}_\text{LLM}$ is the LLM backbone and  $\mathcal{M}_\text{F}$ is the feature-extracting audio tower for embedding the input audio as audio tokens.
We perform our untargeted attack by perturbing the input audio such that the audio tokens differ significantly after the addition of the adversarial noise.
This lets our attack only target the audio tower of the model, making it much faster and more efficient, rather than optimizing with respect to the entire LLM.
Formally, we want to add a stealthy adversarial noise $\delta \in [-\epsilon, \epsilon]^L$ to a benign speech audio $x \in [-1,1]^L$ such that $\mathcal{M}_\text{F}(x)$ and $\mathcal{M}_\text{F}(x+\delta)$ are farther in terms of $\ell_2$ distance.  
We can mathematically write this threat model as:
\begin{align*}
\max_\delta \|\mathcal{M}_\text{F}(x+\delta) - \mathcal{M}_\text{F}(x) \|_2 \quad s.t. \quad \|\delta\|_\infty \leq \epsilon
\end{align*}
where $\epsilon$ denotes the maximum amplitude of the adversarial audio.
To facilitate the optimization process, we can rewrite the objective using a loss function defined as:
\begin{align}
\mathcal{L}(\delta) = -\|\mathcal{M}_\text{F}(x+\delta) - \mathcal{M}_\text{F}(x) \|_2^2
\label{eq:untargeted-loss}
\end{align}
In order to find a viable solution, we can iteratively optimize for an adversarial audio using gradient descent with the following update rules:
\begin{align*}
\hat{\delta}_l &= \delta_{l-1} - \alpha \nabla_{\delta=\delta_{l-1}} \mathcal{L}(\delta), \\
\delta_l^\text{clip} &= clip(\hat{\delta}, -\epsilon, \epsilon), \quad \text{and} \\
\delta_l &= clip(x + \delta_l^\text{clip}, -1, 1) - x
\end{align*}
where $\alpha$ is the learning rate and $clip$ ensures each of the elements in the tensor maintains its valid range.
The audio features $\mathcal{M}_\text{F}(x)$ generated by Qwen2-Audio model is a high-dimensional tensor belonging to $\mathbb{R}^{750\times 1280}$. 
This can lead to the optimization focusing on minimizing the $\ell_2$ distance between only some of the coordinates of the flattened audio feature vector.
In order to tackle this, we encourage the optimization to minimize all the coordinates evenly using randomized masks at every iteration step.
That is, the loss objective in Equation~\ref{eq:untargeted-loss} is modified to be:
\begin{align*}
&\mathcal{L}(\delta) = -\|M \odot \mathcal{M}_\text{F}(x+\delta) - M \odot \mathcal{M}_\text{F}(x) \|_2^2,\\
&s.t.\quad M \sim B(\text{dim}(\mathcal{M}_\text{F}), 1/2)
\end{align*}
where $M$ is a binary mask sampled from a Bernoulli distribution of dimension same as the dimension of the output of $\mathcal{M}_\text{F}(\cdot)$.

\textbf{Target design.}
Our goal is to craft adversarial noises that can affect the utility of ALLMs, say an AI voice agent, by degrading their automatic speech recognition capabilities.
We plan to add our adversarial noises to benign speech commands of innocent users, such that the model does not recognize the original input speech command of the user.
This can practically lead to users not relying on the input speech feature of voice AI assistants.

\textbf{Experimental details and evaluations.}
We use the Qwen2-Audio model as our target ALLM.
For our experiments, we sample benign speech signal $x$ from the LibriSpeech dataset~\citep{panayotov2015librispeech}.
We perform the optimization to find adversarial noises with a batch of 100 four-second clips from the LibriSpeech.
Similar to the targeted attacks, the learning rate is set as 0.0002 to optimize for a mono-channel 16 kHz adversarial noise $\delta$ with $\epsilon$ set as either 0.01 or 0.1.
For evaluations, we look at 2000 four-second-long speech samples from LibriSpeech.
Since we are evaluating the attack's effectiveness in degrading the model's speech recognition ability, the model is given an input text prompt $s=$ `Only generate transcript in English.'.

When evaluating the model's performance in the presence of adversarial noise, the input consists of the text prompt $s$ and the modified audio signal $x+\delta$. 
Here, $x$ represents the clean speech signal from LibriSpeech, and $\delta$ is the carefully crafted, stealthy noise generated by our optimization method.
As a point of comparison, we establish a random baseline. In this baseline condition, the model's performance is evaluated using the same text prompt $s$ but with an audio input of $x+\delta_r$. 
The term $\delta_r$ signifies uniform random noise, with its magnitude intentionally set to be comparable to that of the optimized adversarial noise $\delta$.

To quantify the effectiveness of our attack, we employ three primary metrics: Word Error Rate (WER), Perplexity (PPL), and Attack Success Rate (ASR).
The WER metric measures the discrepancy in words between the transcript generated by the model for an audio input (say, random or adversarial) and the transcript generated for the corresponding clean, benign audio input.
PPL, as defined in Equation~\ref{eq:ppl}, quantifies the model's uncertainty or surprise when generating a specific sequence of output text tokens $t \in |V|^n$, given an input audio $x$ and the text prompt $s$. 
For example, if the model is input with an adversarial audio sample $x+\delta$ and text prompt $s$ to obtain an output text $t$.
We would expect the PPL of the text $t$ with respect to the inputs $s$ and $x+\delta$ to be high if the attack is effective.

\begin{figure*}[htb]
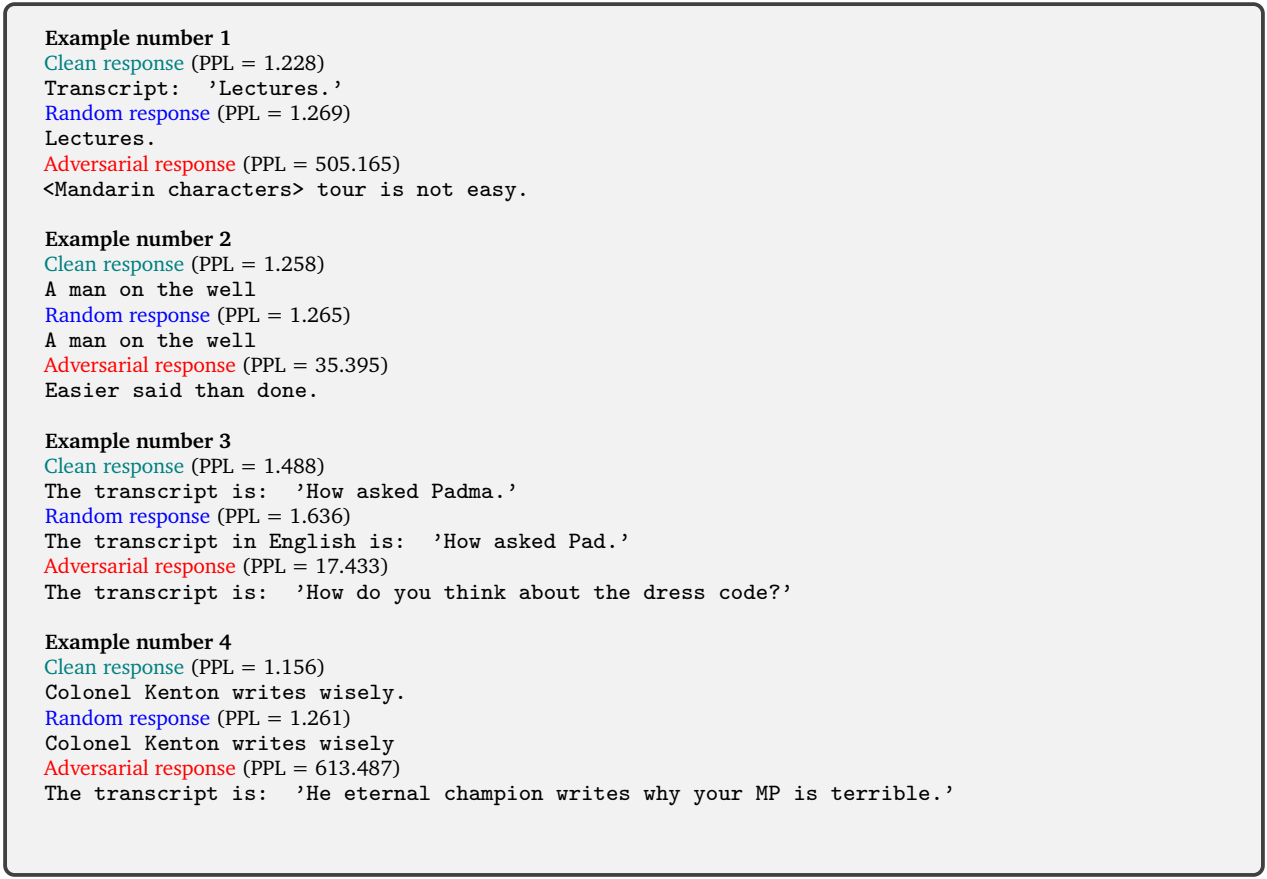

\begin{tcolorbox}[enhanced]
\scriptsize
\textbf{Example number 1}\\
\textcolor{teal}{Clean response} (PPL = 1.228)\\\texttt{Transcript: 'Lectures.'}\\
\textcolor{blue}{Random response} (PPL = 1.269)\\\texttt{Lectures.}\\
\textcolor{red}{Adversarial response} (PPL = 505.165)\\\texttt{<Mandarin characters> tour is not easy.}\\\\
\textbf{Example number 2}\\
\textcolor{teal}{Clean response} (PPL = 1.258)\\\texttt{A man on the well}\\
\textcolor{blue}{Random response} (PPL = 1.265)\\\texttt{A man on the well}\\
\textcolor{red}{Adversarial response} (PPL = 35.395)\\\texttt{Easier said than done.}\\\\
\textbf{Example number 3}\\
\textcolor{teal}{Clean response} (PPL = 1.488)\\\texttt{The transcript is: 'How asked Padma.'}\\
\textcolor{blue}{Random response} (PPL = 1.636)\\\texttt{The transcript in English is: 'How asked Pad.'}\\
\textcolor{red}{Adversarial response} (PPL = 17.433)\\\texttt{The transcript is: 'How do you think about the dress code?'}\\\\
\textbf{Example number 4}\\
\textcolor{teal}{Clean response} (PPL = 1.156)\\\texttt{Colonel Kenton writes wisely.}\\
\textcolor{blue}{Random response} (PPL = 1.261)\\\texttt{Colonel Kenton writes wisely}\\
\textcolor{red}{Adversarial response} (PPL = 613.487)\\\texttt{The transcript is: 'He eternal champion writes why your MP is terrible.'}\\\\
\label{colorbox:untargeted}
\end{tcolorbox}

\caption{Examples of successful untargeted attacks comparing outcomes of the ALLM to clean speech signals, and the corresponding signals with random and adversarial noises in the background.
The significantly higher perplexity scores of our adversarial examples than those of random examples indicate our attack effectiveness.}
\label{fig:untargeted_examples}
\end{figure*}

We consider an attack to be successful if the PPL of the output text generated from an adversarially perturbed audio input exceeds a predetermined threshold. 
For example, ASR@99\% calculates the proportion of adversarial audio inputs ($x+\delta$) that result in a PPL value greater than the 99$^{\text{th}}$ percentile of PPL values obtained when the model processes audio inputs with no added noise.
The results are shown in Table~\ref{tab:additive_noise_impact}.
As shown in the table, both random and adversarial noises affect the model performance.
However, the adversarial setting leads to much extreme degradation in the model's speech recognition performance.
We show some of the successful attacks in Fig.~\ref{fig:untargeted_examples}.

\begin{table*}[t]
    \centering
    \caption{PPL, WER (mean $\pm$ standard deviation), and ASR computed for various input noises with the Qwen2-Audio-Instruct model and the LibriSpeech test dataset. }
    \label{tab:additive_noise_impact}
    \begin{tabular}{@{}c|cccc@{}}
        \toprule
        \textbf{Additive Noise} & \textbf{WER}    & \textbf{PPL}      & \textbf{ASR@99\%} & \textbf{ASR@95\%} \\
        \midrule
        No Noise                & -               & 1.16 $\pm$ 0.15   & -                 & -                 \\ \midrule
        Random ($\epsilon=0.01$)& 0.21 $\pm$ 1.09 & 1.25 $\pm$ 0.38   & 5.65\%            & 12.35\%           \\
        Adversarial ($\epsilon=0.01$)& 0.28 $\pm$ 1.15 & 1.81 $\pm$ 11.37  & 15.45\%           & 32.95\%           \\ \midrule
        Random ($\epsilon=0.1$) & 0.21 $\pm$ 0.85 & 1.35 $\pm$ 0.82   & 17.15\%           & 30.95\%           \\
        Adversarial ($\epsilon=0.1$)& 0.55 $\pm$ 1.69 & 10.26 $\pm$ 34.04 & 62.75\%           & 70.15\%           \\
        \bottomrule
    \end{tabular}
\end{table*}

\section{Scale the Attacks to the Real World}
\label{sec:jailbreak-real-world}

This section addresses the feasibility of transferring our adversarial attacks on audio-based LLMs to real-world scenarios.
Specifically, we investigate whether adversarial audio samples, when played by an adversary through a speaker, can manipulate the output of an audio-based LLM operating on an innocent user's device. 
A key challenge lies in the direct transfer of adversarial perturbations optimized using Equations~\ref{eq:targeted1},~\ref{eq:targeted2},~and~\ref{eq:untargeted-loss} to practical settings.

To mitigate this issue, we propose optimizing adversarial audio samples by incorporating various audio augmentation techniques. 
Given a composition of $m$ distinct augmentation functions $\mathcal{A} = \mathcal{A}_1 \circ \ldots \circ \mathcal{A}_m$, the gradient descent step for the attack is modified as follows:
\begin{align*}
    \hat{x}_l &= x_{l-1} - \alpha \nabla_{x=x_{l-1}} \mathcal{L}(\mathcal{A}(x))
\end{align*}
We conduct experiments employing three audio augmentation techniques: translation, additive noise, and SpecAugment~\citep{park2019specaugment}. Each of these techniques is detailed below.

\textbf{Translation.} The purpose of translation augmentation is to enhance the robustness of our adversarial audio samples against temporal shifts that may occur during the recording or playback process in physical environments. This augmentation can be mathematically represented as:
\[
\mathcal{A}_{\text{translation}}(x) = x_{i:L} \oplus x_{1:i},~s.t.~i \sim \mathcal{U}[1, L]
\]
Here, $\oplus$ denotes vector concatenation and $\mathcal{U}$ denotes a uniform distribution.

\textbf{Additive noise.} To improve the resilience of our attack against background noises present in real-world settings, we incorporate uniform noise during the adversarial optimization process. This augmentation is formulated as:
\begin{align*}
&\mathcal{A}_{\text{noise}}(x) = clip(x + r, -1, 1),\\
&s.t.\quad r \sim \mathcal{U}[-\epsilon_{\text{noise}}, \epsilon_{\text{noise}}]
\end{align*}

In this formulation, $\epsilon_{\text{noise}}$ represents the maximum amplitude of the randomly generated additive noise. Following the addition of noise, the audio signal is clipped to ensure that its amplitude remains within the valid range of -1 to 1.

\textbf{SpecAugment.}
We modify the adversarial audio in the frequency domain to make our attack robust to spectral perturbations that might occur when the audio is recorded by a user's device.
SpecAugment performs this by randomly masking up to $n\_mask$ frequency bands, each of up to width $n\_size$.
Let $S = Spec(x) \in \mathbb{R}^{F \times D}$ denote the spectrogram of $x$ while $F$ and $D$ denote the frequency and time axes, respectively.
$M(S) = S'$ is produced by masking the spectrogram randomly along the frequency axis.
The masked spectrogram is then used to reconstruct the audio by inverse STFT as $x' = InvSpec_x(S')$.
The resulting audio is then rescaled to match its maximum amplitude with the original audio $x$.
The SpecAug technique can be written as:
\begin{align*}
&\mathcal{A}_{\text{spec}}(x) = Rescale_x \circ InvSpec_x \circ M \circ Spec (x),\\
&s.t.~Rescale_x(x') = x'/\max(|x'|) \cdot \max(|x|)
\end{align*}

\subsection{Real World Targeted Attacks}

\begin{table*}[h]
    \centering
    \caption{Ablation study showing the effect of various audio augmentation techniques on the real-world model attack optimization.}
    \label{tab:augmentation-effect-results}
    \begin{tabular}{@{}c|c|ccccc@{}}
        \toprule
        \multirow{3}{*}{\textbf{Augmentations}} & {Translation} & X & $\surd$ & $\surd$ & $\surd$ & $\surd$ \\
        & {Additive noise} & X & X & $\surd$ & X & $\surd$ \\
        & {SpecAugment} & X & X & X & $\surd$ & $\surd$ \\
        \midrule
        \multicolumn{2}{c|}{\textbf{Accuracy}} & 0\% & 0\% & 0\% & 70\% & 100\% \\
        \bottomrule
    \end{tabular}
\end{table*}

\textbf{Target design.}
We want a scenario where the attacker is playing their adversarial noise with their sound source through the air.
The attacker's goal is to hack an innocent user's device that uses an audio-based LLM, say a voice AI assistant. 
For example, the attacker might play an adversarial audio that may trigger the innocent user's AI assistant to wake up and follow a harmful command such as ``Hey Qwen, send money to X''.

\textbf{Experimental details and evaluations.
}
We perform our experiments on the Qwen2-Audio model with 5000 attack iterations.
We use the following default hyperparameters for the attack:
$\alpha$ of 0.0002, $\epsilon$ of 0.1, audio duration of 4 seconds, mini-batch size of 20, $\epsilon_\text{noise}$ of 0.02, $n\_mask$ of 10, and $n\_size$ of 50.
After we optimize the targeted adversarial noise, we play the optimized audio through air using the speaker of an HP Chromebook to simulate the adversary.
In order to simulate an innocent user, we record the audio played by the adversary using an iPhone 15.
The recorded audio is then input into the Qwen2-Audio model to generate text output.
We perform string matching similar to the previous section to check for the presence of the target output $t_{1:n}$ in generated text output.
We evaluate the attack success accuracy with generated outputs over 10 different seeds.

In Table~\ref{tab:augmentation-effect-results}, we show the effectiveness of each of the audio augmentation techniques for the real-world jailbreaking experiments.
As shown in the table, no augmentations give us zero attack success.
The experimental settings with SpecAugment show effectiveness in obtaining non-zero attack success.
This shows that SpecAugment is the key augmentation technique contributing to our attack success.
Additive noise augmentation helps in boosting the attack success from 70\% to 100\% with the presence of SpecAugment.
These attacks show the potential of adversaries to invoke voice-based AI assistants and make them perform harmful tasks in the real world, scalable to multiple innocent users.

\subsection{Real World Untargeted Attacks}

\textbf{Target design.}
Our objective as the attacker is to play an adversarial noise through a sound source, such as a speaker, through the air, such that an innocent user using their audio-based LLM would get a worse speech recognition performance.
The adversarial noise recorded along with the user's speech signal would result in the audio-based LLM performing worse, and hence making these models unreliable to other users.

\textbf{Experimental details and evaluations.}
We perform our experiments on the Qwen2-Audio model with 5000 attack iterations. 
We use the following default hyperparameters for the attack: $\alpha$ of 0.0002, $\epsilon$ of 0.1, audio duration of 4 seconds, mini-batch size of 100, $\epsilon_\text{noise}$ of 0.02, $n\_mask$ of 10 and $n\_size$ of 50.
After we optimize the untargeted adversarial noise, we play the optimized audio through the air using the speaker of an HP Chromebook to simulate the adversary along with clean four-second-long audio samples from the LibriSpeech dataset.
We record 25 different LibriSpeech samples.
In order to simulate an innocent user, we record the audio played by the adversary using an iPhone 15.
The recorded audio is then input into the Qwen2-Audio model to generate text output.
We evaluate outputs generated with four different seeds, leading to a total of 100 generations.
We also record baseline audios without any adversarial noise and with uniform noise of similar magnitudes to the adversarial noise.

\begin{figure*}[htb]
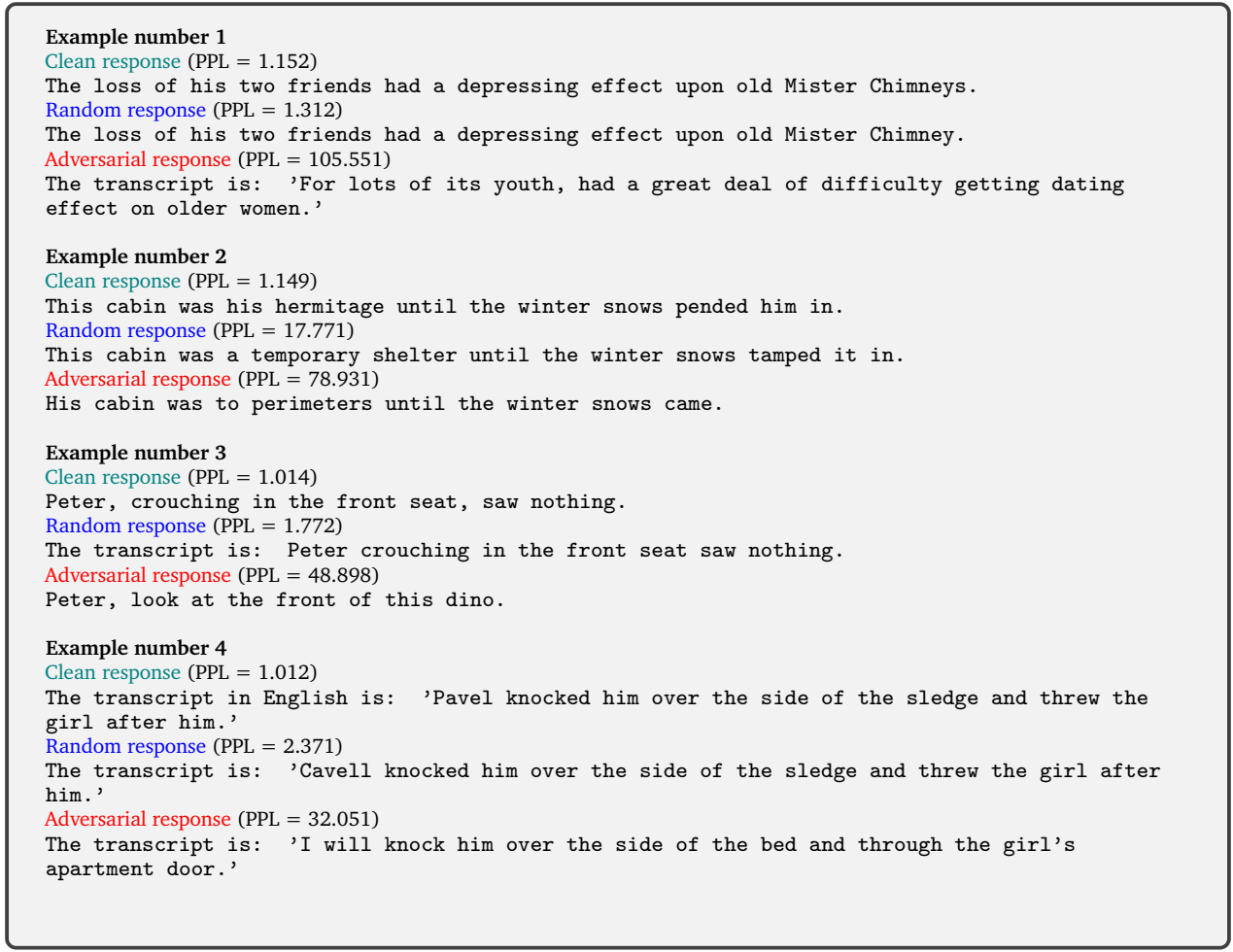

\begin{tcolorbox}[enhanced]
\scriptsize
\textbf{Example number 1}\\
\textcolor{teal}{Clean response} (PPL = 1.152)\\\texttt{The loss of his two friends had a depressing effect upon old Mister Chimneys.}\\
\textcolor{blue}{Random response} (PPL = 1.312)\\\texttt{The loss of his two friends had a depressing effect upon old Mister Chimney.}\\
\textcolor{red}{Adversarial response} (PPL = 105.551)\\\texttt{The transcript is: 'For lots of its youth, had a great deal of difficulty getting dating effect on older women.'}\\\\
\textbf{Example number 2}\\
\textcolor{teal}{Clean response} (PPL = 1.149)\\\texttt{This cabin was his hermitage until the winter snows pended him in.}\\
\textcolor{blue}{Random response} (PPL = 17.771)\\\texttt{This cabin was a temporary shelter until the winter snows tamped it in.}\\
\textcolor{red}{Adversarial response} (PPL = 78.931)\\\texttt{His cabin was to perimeters until the winter snows came.}\\\\
\textbf{Example number 3}\\
\textcolor{teal}{Clean response} (PPL = 1.014)\\\texttt{Peter, crouching in the front seat, saw nothing.}\\
\textcolor{blue}{Random response} (PPL = 1.772)\\\texttt{The transcript is: Peter crouching in the front seat saw nothing.}\\
\textcolor{red}{Adversarial response} (PPL = 48.898)\\\texttt{Peter, look at the front of this dino.}\\\\
\textbf{Example number 4}\\
\textcolor{teal}{Clean response} (PPL = 1.012)\\\texttt{The transcript in English is: 'Pavel knocked him over the side of the sledge and threw the girl after him.'}\\
\textcolor{blue}{Random response} (PPL = 2.371)\\\texttt{The transcript is: 'Cavell knocked him over the side of the sledge and threw the girl after him.'}\\
\textcolor{red}{Adversarial response} (PPL = 32.051)\\\texttt{The transcript is: 'I will knock him over the side of the bed and through the girl's apartment door.'}\\\\
\label{colorbox:untargeted-real-world}
\end{tcolorbox}
\caption{Examples of successful instances when our untargeted attacks work in the real-world. 
Here, adversarial noises played by the attacker's Chromebook significantly affect the speech transcription performance of an innocent user's ALLM recording speech through their iPhone when compared to the effect of random background noises.}
\label{fig:untargeted_examples_rw}
\end{figure*}

We show example real-world samples in Fig.~\ref{fig:untargeted_examples_rw}.
As shown, the perplexity statistics are much worse when the adversarial audio is played compared to when the uniform noise is played.
These examples show that our untargeted attacks can be transferred to real-world scenarios, affecting the utility of audio-based LLMs.

\section{Discussion}
\label{sec:discussion}

\subsection{Transferability across System Instructions}
\label{sec:transfer}

OpenAI introduced the GPT Store, a library of GPT models majorly by cutomized using system instructions~\citep{gptstore}.
Here an important question is: ``Can an attack designed for the base model transfer to the same model but with a different system instruction?''.
In this section, we evaluate the transferability of our attacks when we deploy them to models with a different system instruction.
In previous sections, we considered the Qwen2-Audio model with its default system instruction ``You are a helpful assistant''.
In this section, we use the system instruction below which also instructs the model to ignore background noises to make it more resilient to adversarial noises.

\begin{tcolorbox}[enhanced]
\scriptsize
Your name is `Qwen'. You are an AI voice assistant agent capable of performing a lot of actions for the user. You are supposed to listen to the user command and enact accordingly. The user will ask you some questions or to perform some action. Be a capable voice assistant. Ignore background noise in the input audios.
\end{tcolorbox}

We evaluate the real world attacks in \S\ref{sec:jailbreak-real-world} with this setting.
For untargeted attacks, we find that the perplexity statistic with the new system instruction (and with the default system instruction) to be $1.09\pm0.07$ ($1\pm0.09$), $2.54\pm3.49$ ($2.99\pm4.35$), and $12\pm18.72$ ($14.08\pm17.36$), respectively, for clean, random, and adversarial settings.
Though the perplexity scores decrease a bit with the new system instruction for the adversarial setting, we still find the values to be much worse when compared to the random setting.
This shows that our attack can transfer to a new model with a custom system instruction.
For targeted attacks, we find that with both the system instructions, we obtain a 100\% attack success to make the model produce the target output ``Hey Qwen''.
\subsection{Potential Defenses}
\label{sec:defenses}

In this section, we test how our targeted attacks in Section~\ref{sec:jailbreak-real-world} would be affected with input audio augmentation techniques.
We specifically look into audio augmentations which were not employed in the attack optimization process such as audio sample rate modification, noise reduction with spectral gating, and EnCodec compression \citep{encodec}.

We perform our experiments using the adversarial noise we optimized for generating the target string ``Hey Qwen'' from Section~\ref{sec:jailbreak-real-world}.
In Table~\ref{tab:samplerate_effect}, we show our attack success rates when we modify the sample rate of the audio by a rescale factor of 0.4 to 1.4. As we see in the default case, when the rescale factor is 1 or the sample rate of augmented audio is 16 kHz, the attack obtains 100\% success. 
For smaller modifications, that is for rescale factors of 0.8 and 1.2, we find that the original adversarial noise breaks the model 100\% of the time.
However, we find that the recorded adversarial audio with iPhone is sensitive to this perturbation and obtains 0\% success for any perturbation of the sample rate.
In Table~\ref{tab:noise_gate}, we show the results with spectral noise gating.
We use the \texttt{noisereduce} Python package \citep{tim_sainburg_2019_3243139} for noise reduction.
With a default of 100\% noise reduction, the attack succeeds only 0\% with the recorded adversarial audio while with the original adversarial audio it works for 70\% of the time.
For all the other noise reduction parameters, we find the defense fails.
Table~\ref{tab:encodec} shows the results of the attack performance after the adversarial audios are encoded and decoded by the neural codec model EnCodec \citep{encodec}.
EnCodec shows to be the most effective defense technique against our attack defending almost 100\% of the time with varying compression bandwidths.

These findings indicate that the adversarial attacks we developed can be countered in real-world scenarios by applying simple input audio augmentations.
However, if an attacker is aware of these defense mechanisms, they can create adaptive attacks specifically designed to bypass these protective measures.
This idea is supported by one of our experimental observations: adversarial audio created without considering translation augmentation during its optimization phase is not effective when translations are applied during the testing phase.
Conversely, we found that when we include translation augmentations directly into our attack optimization process, the resulting adversarial attacks become resilient and effective even when translations are applied.

\begin{table*}[h!]
    \centering
    \caption{Attack success rate (\%) after sample rate change.}
    \label{tab:samplerate_effect}
    \begin{tabular}{@{}c|cccccc@{}}
        \toprule
        \textbf{Sample Rate Change} & 0.4$\times$ & 0.6$\times$ & 0.8$\times$ & 1.0$\times$ & 1.2$\times$ & 1.4$\times$ \\
        \midrule
        \textbf{Original Audio}     & 0   & 0   & 100 & 100 & 100 & 0   \\
        \textbf{Recorded Audio}     & 0   & 0   & 0   & 100 & 0   & 0   \\
        \bottomrule
    \end{tabular}
\end{table*}

\begin{table*}[h!]
    \centering
    \caption{Attack success rate (\%) after noise reduction using spectral gating.}
    \label{tab:noise_gate}
    \begin{tabular}{@{}c|ccccc@{}}
        \toprule
        \textbf{Noise Reduction (\%)} & 100 & 75  & 50  & 25  & 0   \\
        \midrule
        \textbf{Original Audio}       & 70   & 100 & 100 & 100 & 100  \\
        \textbf{Recorded Audio}       & 0  & 100 & 100 & 100 & 100 \\
        \bottomrule
    \end{tabular}
\end{table*}

\begin{table}[h!]
    \centering
    \caption{Attack success rate (\%) across different EnCodec compression bandwidth levels.}
    \label{tab:encodec}
    \begin{tabular}{@{}c|ccccc@{}}
        \toprule
        \textbf{Bandwidth (kbps)} & 1.5 & 3 & 6 & 12 & 24 \\
        \midrule
        \textbf{Original Audio}   & 0   & 0 & 0 & 0  & 50 \\
        \textbf{Recorded Audio}   & 0   & 0 & 0 & 0  & 0  \\
        \bottomrule
    \end{tabular}
\end{table}
\section{Conclusion}

This paper demonstrates that ALLMs possess a significant vulnerability to adversarial audio inputs. Our findings show that an attacker with white-box access can craft stealthy audio perturbations to manipulate these models. We illustrate two primary attack vectors---targeted attacks that compel the model to perform specific, potentially malicious actions, and untargeted attacks that degrade the model's response quality and utility.

A crucial contribution of this work is demonstrating the feasibility of these attacks in real-world scenarios. By incorporating audio augmentation techniques like SpecAugment into the optimization process, we created adversarial audio robust enough to be played over the air, successfully manipulating an ALLM on an innocent user's device. These attacks proved transferable, remaining effective even when system instructions were changed to ignore background noise.

While we explored potential defenses and found that neural audio compression like EnCodec can be highly effective, we also note that attackers could likely adapt to circumvent such measures. These findings highlight a critical security risk, urging caution in the open-sourcing of audio-based models and underscoring the necessity of developing more robust, adaptive defenses.

\bibliography{ref}

\end{document}